\documentstyle[12pt,aasms4]{article}
\def\alwaysmath#1{\ifmmode{#1}\else{$#1$}\fi}

\def\msun{\alwaysmath{\,M_\odot}}

\def\morethan#1{{ et al.}}  
\def\simlt{\lower.5ex\hbox{$\; \buildrel < \over \sim \;$}}
\def\simgt{\lower.5ex\hbox{$\; \buildrel > \over \sim \;$}}

\slugcomment{\footnotesize In press to ApJ.
}

\lefthead{Ventura et al.}
\righthead{Selfpollution on Globular Cluster stars}

\begin{document}

 \title{Predictions for selfpollution in Globular Cluster Stars}
\author{ Paolo
Ventura\altaffilmark{1}, Francesca D'Antona\altaffilmark{1} }
\affil{Osservatorio Astronomico di Roma, Via Frascati 33,
00040 Monte Porzio Catone, Italy}

\author{Italo Mazzitelli\altaffilmark{2}}
\affil{Istituto di Astrofisica Spaziale del CNR, Via del Fosso del Cavaliere
100, 00133  Rome, Italy}
\author{Raffaele Gratton\altaffilmark{3}}
\affil{Osservatorio Astronomico di Padova, Vicolo dell'Osservatorio 5, 35122
Padova, Italy}
\altaffiltext{1}{email: dantona, paolo
@coma.mporzio.astro.it}
\altaffiltext{2}{email: aton@hyperion.ias.rm.cnr.it}
\altaffiltext{3}{email: gratton@pd.astro.it}

\begin{abstract}

Fully evolutionary models have been built to follow the phases of
Asymptotic Giant Branch (AGB) evolution with mass loss, for metal mass
fractions from Z$= 2 \times 10^{-4}$\ to Z$= 4 \times 10^{-3}$.  The Hot
Bottom Burning (HBB) at the base of the convective envelope is followed by
fully coupling nuclear burning and non instantaneous mixing. The models also
show the occurrence of a spontaneous (not induced by overshooting) third
dredge-up. For the first time we find that temperatures close or even larger
than $10^8$K are achieved at low Z: the full CNO cycle operates
at the basis of the envelope, the $^{16}O$\ abundance for the most metal poor
models of mass 4 and 5\msun\ is drastically reduced, and sodium and
aluminium production by proton capture on neon and magnesium can occur.
Lithium is first largely produced in the envelope and then burned completely,
so the average lithium abundance in the expelled envelope is a factor up to
5 times smaller than the initial one, but it is never completely depleted.

These results may be relevant for the evolution of primordial massive
Globular Clusters: we suggest that the low mass stars may have been polluted
at the surface by accretion from the gas lost from the evolving intermediate
mass stars at early ages (1-2$\times 10^8$yr). In this hypothesis, we should
expect that the polluted stars show smaller abundance of oxygen, larger
abundances of products of advanced nucleosynthesis as Na and Al, and lower,
but never negligible, abundances of lithium. The abundance spreads should be
smaller in clusters of higher metallicities, where the lithium in the
polluted stars could be larger than in the non polluted stars. \end{abstract}

\section{Introduction}
In recent years, determination of chemical abundances in Globular Cluster
stars has confirmed the inhomogeneity of their surface composition for what
concerns the abundances of the light elements from lithium to aluminium (see,
for reviews, Smith 1987 and Kraft 1994). In particular, red giants in some
GCs show a Na--O anticorrelation (e.g. Kraft et al. 1997) which seems the
clear sign of `in situ' processes, by which some still unexplained deep
mixing mechanism brings at the stellar surface the products of complete CNO
cycling occurring deep in the stellar interiors (e.g. Langer, Hoffman \&
Sneden 1993).
This hypothesis is until now not confirmed by computation of adequate stellar
models. In addition, the deep mixing should take place {\it only} in GC
stars, as the analogous stars in the field are much more homogeneous in
abundances.

Also the lithium abundances at the Turnoff (TO) of some GCs show spreads
which are larger than the spread among similar halo stars (King, Stephens
\& Boesgaard 1998; \cite{Pasq}), 
which, in the range of GCs metallicities, are known to be very
homogeneous ($\log N(Li)=2.24 \pm 0.05$, \cite{BM97}; see also Ryan,
Norris \& Beers 1999), possibly indicating that we are witnessing 
the abundance left by the Big Bang nucleosynthesis. Of
course, any lithium spread can not be linked with simplicity to advanced
nucleosynthesis, being lithium a very fragile element, which would be
completely destroyed at the temperature necessary to produce sodium. In
addition, for any spread of abundances at the TO, it is not possible to
invoke `in situ' mechanisms, apart from gravitational settling and thermal
diffusion (e.g. \cite{chab92}). In this context, Gratton et al. (2001) have
obtained recently very important results: the TO and subgiant stars in
the GC NGC 6752 show an oxygen spread anticorrelated with sodium, and a
Mg-Al anticorrelation, which certainly can not be
explained by any `in situ' process.

A different model for the GC stars inhomogeneities, not necessarily
alternative with the `in situ' scenario (see, e.g. Denissenkov et al. 1998),
has been put forward in the eighties: the `selpollution' or `self-enrichment'
hypothesis. The idea is that the field stars and the GC stars are born in a
very different environment which plays a role in their evolution. The surface
composition of the GC stars which we see today evolving, in fact, may have
been contaminated by accretion from the ejecta of more massive stars which
have evolved during the past history of the cluster. Cottrell \& Da
Costa (1981) suggested that the observed sodium and aluminium variations were
the results of reactions occurred in an early generation of intermediate
mass stars. These stars are also the best candidates to produce accretion 
{\it on the already formed stars}. In fact, we would not expect accretion on
stars from possible supernova ejecta; in addition, we already know that today
there is practically no gas or dust in the center of GCs, probably due to the
strong UV flux from blue Horizontal Branch and post-AGB stars, which helps in
removing the mass lost from red giants (\cite{FF77}).

However, the intermediate mass stars (M $\simeq 3 - 6$\msun) 
lose mass by low velocity
stellar winds, which can remain into the cluster, if it is massive enough,
and concentrate in the core. These stars, in addition, suffer a very fast
evolution through the hot Planetary Nebulae region, before becoming white
dwarfs (\cite{WF86}, Vassiliadis \& Wood 1994), so that their input UV
energy is not enough to expell the gas. In these conditions, the low mass
stars, passing through the central regions of the GC, may accrete it in
appreciable quantities. This hypothesis was put forward first by D'Antona,
Gratton \& Chieffi (1983).
Gratton (2001) has recently elaborated on this idea. An
additional important feature of such a model is that the massive AGB stars
have a very interesting envelope nucleosynthesis, as the basis of their
convective envelopes becomes very hot during the evolution: they suffer `Hot
Bottom Burning' (HBB), which mainly produces -and burns- lithium, and cycles
the CNO elements.

The interesting novelty in the self-pollution scenario is that today we are
able to {\it predict} the main composition of the ejecta of the intermediate
mass AGBs, as a function of the metallicity, by means of new, sophisticated
models. We can in fact model in detail the Thermal Pulse phase, and have
developped a code which can follow selfconsistently the chemical mixing and
nuclear burning in the envelope. Further, we have been able to calibrate the
mass loss during the AGB phase, based on observational properties of such AGB
stars in the Magellanic Clouds. Before embarking in a detailed study of the
modalities of the gas dynamics and of the accretion process, we
first compute the nucleosynthesis expected in these stars, and therefore the
possible main elements and isotopic ratios variations as a function of the
initial mass and metallicity. We follow the evolution until the envelope mass
is reduced to a relatively small fraction of the initial mass, and the models
are reasonably close to the planetary nebula phase.

We have specifically computed the evolution of lithium, carbon isotopes,
nitrogen and oxygen isotopes for intermediate mass stars (3$\leq M/M_\odot
\leq 6.5$), having metals mass fractions from Z$= 2 \times 10^{-4}$\ to
$Z=0.01$. Full description of the results and the elements yields will
be given in a forthcoming paper.
We describe our models in Section 2, their input physics and important
differences with previous computations.
Section 3 shortly presents the results and Section 4 summarizes the possible
main observational predictions.

\section{The models}
We have developped new models of TP AGB stars in the phase of Hot Bottom
Burning (Mazzitelli, D'Antona \& Ventura 1999). For a detailed description 
of the input physics see also Ventura, D'Antona \& Mazzitelli (2000). 
In these models we follow in detail the
nucleosynthesis of 14 elements ($^1H$, $^2D$, $^3He$, $^4He$, $^7Li$, $^7Be$,
$^{12}C$, $^{13}C$, $^{14}N$, $^{15}N$, $^{16}O$, $^{17}O$, $^{18}O$,
$^{22}Ne$) and 22 associated nuclear reactions. We do not follow therefore 
the detailed nucleosynthesis of trace elements, but the decrease in
the $^{16}O$\ abundance, and the increase of $^{17}O$\ monitor when the
complete CNO Cycle is active. {\bf For the phases of Oxygen cycling, the
physical conditions at the envelopes bottoms are such that they allow
production of sodium by burning of the original $^{20}Ne$ and $^{22}Ne$,
and of aluminium through burning of the magnesium isotopes (e.g. Langer
et al. 1993), but we cannot give quantitative estimates for the expected
anticorrelations O-Na and Mg-Al. We are now planning computations including
a network of 30 isotopes to solve the problem selfconsistently.} 

The detailed lithium nucleosynthesis is
followed directly in the evolutionary models, in which first the
lithium remnant from the previous possible dilution phases is destroyed,
then fresh lithium is produced through the reaction $^3He+^4He \rightarrow
^7Be \rightarrow ^7Li$ (the `Cameron Fowler' mechanism ---Cameron \&
Fowler 1971),
and finally it is burnt completely when the $^3He$\ of the envelope is fully
depleted. To follow selfconsistently the nuclear evolution, we use a scheme
which couples the nuclear burning and chemical mixing in the envelope, as
first included in full stellar models by Sackmann \& Boothroyd (1992).

These are the first models which provide fully evolutionary results of
nucleosynthesis by HBB in AGB for metallicities as low as the
GCs metallicities.
Detailed results for AGBs of Z=0.02 and Z=0.005 have been published by
Forestini \& Charbonnel (1997), 
but they followed explicitly only a few TPs and adopt an
extrapolation for the rest of the evolution. Their models differ from ours in
the choice of the treatment of convection\footnote{We adopt the
`Full Spectrum of Turbulence' (\cite{CM91}) convection, in the formulation by
Canuto, Goldman \& Mazzitelli (1996). 
This model predicts HBB temperatures larger than in the
solar-calibrated Mixing Length model adopted by Forestini \& Charbonnel (1997), 
see D'Antona \& Mazzitelli (1996). 
However, we tested that also MLT models with a solar calibrated
mixing length reach very large HBB
temperatures for the lowest computed metallicities and deplete oxygen.} and
of a mass loss formalism, which probably underestimates
mass loss. Denissenkov et al. (1998) also consider detailed nucleosynthesis
in AGBs as possible source of self-pollution in GCs, but they adopt a
`parametrized nucleosynthesis' approach and do not compute full evolutions.
The yields obtained do not depend sensibly on the mass loss formulation and
on its calibration, apart from the lithium yield. We adopt Bl\"ocker (1995)
schematization for mass loss, and have recently sorted out a way of
calibrating its free parameter on the basis of the Magellanic Clouds
observations (\cite{ventura00}), showing that the lithium yield varies more
or less linearly with the mass loss rate. We make here the hypothesis that
this calibration holds also for lower metallicity stars.

Among the other elements, we follow the carbon and nitrogen evolution.
These abundances are affected by the occurrence of the `third dredge up'
(Iben 1975) of carbon from the helium shell. This process produces, at lower
luminosities, the phenomenon of Carbon stars. The masses we consider here,
due to the HBB, cycle carbon to nitrogen and do not appear as Carbon stars
even if the third dredge up is operating. This process is however still not
well understood: for the solar chemistry, most models, including ours, need a
treatment of overshooting below the formal convective envelope to produce
Carbon stars in the range of luminosity and metallicities for which they
occurr in nature. We have considered models in which Carbon stars are formed,
by allowing a detailed consideration of non-instantaneous mixing by
`overshooting' (Ventura, D'Antona \& Mazzitelli 1999), 
but we did not include this treatment in
the present models, as at low metallicities the models achieve the third
dredge up spontaneously. As we do not include overshooting below the
convective bottom, however, our carbon and nitrogen abundances are to be
considered as {\it lower limits}. {\bf Notice that the surface helium
in these stars is not particularly peculiar: helium dredge-up in the HBB
phase is negligible (contrary to the models in which deep mixing is
artificially enforced to explain `in situ' the peculiarities of GC red 
giants, e.g. Weiss, Denissenkov \& Charbonnel 2000). 
However these stars have suffered the second
dredge up, so the helium content of the ejecta is $Y\sim 0.29$, to be compared
with the initial $Y=0.23$.} Details will be published elsewhere.

\section{Results}
In presenting the results in the figures, we show the physical quantities
along the computed evolutions as a function of the total `mass', which is
decreasing due to mass loss. In this way, we also have an immediate
understanding of how much a phase is important for the chemical yields.
Figure \ref{fig_1} shows the temperature at the bottom of the convective
envelope ($T_{bce}$) along the evolution of models
of 4, 5 and 6\msun\ (Z=0.01, left) and 3, 4 and 5\msun\ (Z=2$\times
10^{-4}$, right).

We notice that $T_{bce}$\ for the same initial mass is larger the lower is
the metallicity. The most massive and metal poor models reach surprisingly
large values of $T_{bce}$, although we do not allow for any kind of
overshooting below the formal convective region. Above $ T_{bce} \simeq 8
\times 10^7$K the CNO cycle is complete, and oxygen becomes depleted.
Fig. \ref{fig_2} shows the evolution of the ratio between the surface oxygen
abundance and the initial value for stars of $5M_{\odot}$ for the
computed metallicities. We see that for Z=0.01 the abundance
when the total mass is reduced to $\sim 2M_\odot$\ is not different.
Depletion is more important for Z=$4 \times 10^{-3}$ and $10^{-3}$. The track
of $2 \times 10^{-4}$\ shows an initial reduction by a factor $\sim 100$,
which in later phases is reduced to a factor $\sim 10$. Although our
computations do not include the nucleosynthesis past oxygen, the drastic
reduction of $^{16}O$\, with the $^{17}O/^{16}O$ increasing up to $\sim
0.17$, indicates that we are in the presence of a very advanced
nucleosynthesis. At these temperature and densities, Na and Al are produced
by proton capture on the neon and magnesium nuclei respectively.

Figure \ref{fig_3} shows the surface lithium as a function of the mass for the
metallicities Z=$4 \times 10^{-3}$\ and $2 \times 10^{-4}$. Assuming that
all these stars start with
an initial abundance equal to the population II abundance
$\log N(Li)=2.2$, there is a short phase in which lithium is
overproduced by a factor up to $\sim 60$, followed by a more prolonged phase
in which it is depleted by a factor $\sim 10^4$. In the computation of the
total yield, however, the phase of production is very important to balance
the following total desctruction. As a final result, lithium is {\it
depleted} with respect to the Big Bang abundance, by a factor 4--5 at the
lowest metallicities. At the lowest Z, the phase of $^7Li$\ production
lasts for a shorter time. In addition, the radii are smaller and the mass
loss rate is lower. As a result, the total lithium depletion in the ejecta
is maximum. In fact, at $Z=4 \times 10^{-3}$\ the expected depletion is by at
most a factor 2.

Figure \ref{fig_4}\ shows the lithium abundance in the ejected envelopes, as
a function of the metallicity, including also the models for Z=0.01 computed in
Ventura et al. (2000). The horizontal line shows the `primordial' abundance.
This should sure apply to the most metal deficient GCs. We see then that for
these clusters we expect that the polluted stars (in which oxygen is
smaller) should have also a smaller lithium abundance.

For more metal rich GCs, such as, e.g., 47 Tuc, which has a metallicity of
$\simeq 4 \times 10^{-3}$, the situation is more complex. The scarce
available data on lithium (see, e.g., the compilation by Romano et al. 1999)
seem to indicate a value somewhat larger than the plateau value ($\sim 2.3 -
2.4$). Of course, the lithium in the halo stars at this metallicity may
differ from the value in a GC. If we assume that the initial lithium at Z=$4
\times 10^{-3}$\ is still close to the Spite plateau, we see that the gas of
the polluting stars is expected to be of a quite similar or scarcely smaller
abundance. Our mass loss calibration must be believed only within
a factor two or three, and this would alter the lithium abundances by the
same factor, by enhancing or reducing the mass loss rate in the phases of
lithium production (\cite{ventura00}). We might then expect in metal rich
massive GCs either a lithium spread similar to that expected for low
metallicity -if the initial lithium is a bit larger than the plateau value-,
or the opposite behaviour: namely that the polluted stars have a lithium
abundance {\it larger} than the initial one.

Notice that the lithium abundance in the ejecta is never very large, not
even
for the largest metallicities. This in fact was one of the results of our
calibration of the mass loss rate ({\cite{ventura00}), and shows that massive
AGB stars are not important contributors to the galactic lithium enrichment
(Romano et al. 2001).

\section{Predictions}
If the self-pollution model is relevant for GCs, we can derive from our
models some predictions:
\begin{itemize}
\item Metal poor GCs should show the largest spreads of abundances, and in
particular a spread in oxygen, anticorrelated with Na, and probably
an anticorrelation Mg--Al also, similar to what would be expected by `in
situ' very deep mixing in the advanced giant branch phases.
\item Contrary to the expectation of the `in situ' nucleosynthesis, these
abundance anomalies are expected to be present also at the main sequence or
on the subgiant branch. The results by Gratton et al. (2001) on the
anticorrelation oxygen--sodium and magnesium--aluminium for turnoff and
subgiant stars in the massive GC NGC 6752 are in agreement with this model.
\item An interesting test to understand whether the oxygen variations are
due to pollution from the envelopes of intermediate mass stars would be the
detection of lithium abundance correlated with the oxygen abundance. The
process should in fact produce a spread in the lithium, of the same order,
or smaller, than the oxygen variations.
Fully CNO cycled stellar matter is characterized by a huge lithium
depletion,
but the envelopes of the polluting stars have also passed through a phase in
which lithium had been extensively {\it produced} by consumption of the
envelope $^3He$, and some of the mass processed in this way has been
lost by wind. The `normal' lithium detection of two turnoff stars in 47Tuc
(\cite{Pasq}), which have significantly different (anticorrelated) CH and CN,
are compatible with this model.
\item The most metal rich clusters should be characterized by smaller
degrees of elemental variations, but some correlation oxygen -- lithium
should remain. In fact, as we can not trust completely our mass loss
calibration, it is well possible that the polluted stars present a lithium
abundance {\it larger} than the initial one.
\end{itemize}

We thank Alessandro Chieffi for useful comments.

\clearpage
\begin{figure}
\caption{We show the temperature at the bottom of the convective envelopes
as a function of the stellar mass, along the evolution of stars of 6, 5 and
4\msun\ (Z=0.01, left panel, from Ventura et al. 2000) 
and 5, 4 and 3.5\msun\ (Z=2$\times 10^{-4}$,
right panel). Notice that $T_{bce}$ is over $10^8$K at the lower
metallicity.} \label{fig_1}
\end{figure}

\begin{figure}
\caption{Logarithm of the Oxygen abundance with respect to the initial 
$^{16}O$ value along the evolution of stars of
5\msun with different (labelled) metallicities. Models with Z=0.01 were taken
from Ventura et al. (2000)}
\label{fig_2}
\end{figure}

\begin{figure}
\caption{Lithium abundance along the evolution, for Z=$2 \times 10^{-4}$\
(right) and Z=$4 \times 10^{-3}$ (left). Notice first the Lithium production
due to HBB, then the drastic Lithium depletion, when the envelope $^3He$\ is
exhausted. {\bf Note the rapid drop of the lithium abundance following
each pulse, due to the decrease of the temperature at the base of the
envelope, which stops lithium production.}} \label{fig_3} \end{figure}

\begin{figure}
\caption{Lithium abundance in the ejected envelopes as function of the
initial mass and metallicity. The maximum Lithium depletion with respect to
the horizontal line, representing the primordial abundance, is by a factor
$\sim 5$ for the 5\msun\ of Z=2 and 6$\times 10^{-4}$.
These abundances depend on the mass loss rate, which has been carefully
calibrated for Z=0.01 (Ventura et al. 2000). Smaller rates can apply to the
more metal poor environments, leading to a more or less proportional
decrease in the ejected abundance.
} \label{fig_4} \end{figure}

\end{document}